\documentclass{amsart}

\usepackage{bm}
\usepackage[citation-order]{amsrefs}

\newcommand{\ed}{\mathrm{d}}
\newcommand{\R}{\mathbb{R}}
\newcommand{\N}{\mathbb{N}}
\DeclareMathOperator{\rank}{rank}
\DeclareMathOperator{\ld}{\pounds}
\newcommand{\Lie}[1]{\ld_{#1}}

\newcommand{\Del}{\bar\nabla}
\DeclareMathOperator{\Nabla}{\bar\nabla}
\newcommand{\cd}[1]{\Nabla_{#1}}

\theoremstyle{plain}
\newtheorem{thm}{Theorem}
\newtheorem{prp}{Proposition}

\theoremstyle{definition}
\newtheorem{dfn}{Definition}
\newtheorem{exm}{Example}

\theoremstyle{remark}

\title{Degenerate Hessian Structures on Radiant Manifolds}
\author{Miguel \'Angel Garc\'ia-Ariza}
\address{Facultad de Ciencias F\'isico Matem\'aticas, Benem\'erita Universidad Aut\'onoma de Puebla, Apartado Postal 1152, 72000 Puebla, Mexico}
\email{magarciaariza@alumnos.fcfm.buap.mx}

\begin{document}

\begin{abstract}
We present a rigorous mathematical treatment of Ruppeiner geometry, by considering degenerate Hessian metrics defined on radiant manifolds. A manifold $M$ is said to be radiant if it is endowed with a symmetric, flat connection $\Del$ and a global vector field $\rho$ whose covariant derivative is the identity mapping. A degenerate Hessian metric on $M$ is a degenerate metric tensor $g$ that can locally be written as the covariant Hessian of a function, called potential. A function on $M$ is said to be extensive if its Lie derivative with respect to $\rho$ is the function itself. We show that the Hessian metrics appearing in equilibrium thermodynamics are necessarily degenerate, owing to the fact that their potentials are extensive (up to an additive constant). Manifolds having degenerate Hessian metrics always contain embedded Hessian submanifolds, which generalize the manifolds defined by constant volume in which Ruppeiner geometry is usually studied. By means of examples, we illustrate that linking scalar curvature to microscopic interactions within a thermodynamic system is inaccurate under this approach. In contrast, thermodynamic critical points seem to arise as geometric singularities.
\end{abstract}

\maketitle

\section{Introduction}
Our study of degenerate Hessian metrics defined on radiant manifolds is motivated by the geometry of equilibrium thermodynamics. As first pointed out by Weinhold \cite{Weinhold:1975aa}, an interesting consequence of the laws of thermodynamics is that the manifold of equilibrium states of any classical thermodynamic system is naturally endowed with a degenerate metric tensor whose components are given by the Hessian matrix of a thermodynamic potential, computed with respect to the \textit{extensive variables} of the system. The aim of this paper is to describe such metrics in a coordinate-free fashion. To this end, we shall work on an affine finite-dimensional manifold $M$, \textit{i. e}, a smooth manifold furnished with a flat, symmetric affine connection $\bar\nabla$. 

The degenerate metric that we mentioned above is a symmetric, positive semi-definite tensor field $g$ of type $(0,2)$ defined on $M$ that satisfies
\begin{equation}\label{eq:Codazzi}
\Del_Xg(Y,Z)=\Del_Yg(X,Z),
\end{equation} 
\noindent for every local vector fields $X$, $Y$, and $Z$ defined on $M$. When $g$ is positive definite, the pair $(\Del, g)$ is a \emph{Hessian structure} on $M$ \cite{Shima:2007aa}. In contrast to the latter, the structure in which we are interested possesses at least a null vector, this is, there exists a global vector $X$ defined on $M$ such that $g(X,Y)=0$, for any local vector field $Y$. For this reason, we shall refer to the pair $(\Del,g)$ as a \textit{degenerate Hessian structure} on $M$. The tensor field $g$ will be called \emph{degenerate Hessian metric}.

The degenerate Hessian structures appearing in equilibrium thermodynamics have an important local attribute. From Eq. \eqref{eq:Codazzi}, it follows that $g$ is locally written as the Hessian (with respect to $\bar\nabla$) of a local smooth function: the \textit{potential} of the degenerate Hessian structure. In thermodynamics, these potentials are \textit{extensive functions}. In elementary terms, this means that they are homogeneous degree-1 functions of a particular set of coordinates. To achieve a global description of this feature, we provide a suitable, coordinate-free definition of extensive functions by means of a \emph{radiant structure} (Definition \ref{dfn:ext}). It turns out that having extensive local potentials is globally translated to certain compatibility property between the Hessian and the radiant structures (Theorem \ref{thm:ex=>dg}). 

When restricted to certain physically-relevant Hessian submanifolds of $M$, degenerate Hessian metrics acquire a central role in the framework of thermodynamic fluctuation theory, owing to Ruppeiner's so-called \emph{interaction hypothesis}. This statement---the core of \emph{Ruppeiner geometry}---asserts that the scalar curvature of this submanifolds is related to critical behavior, and that its sign yields information about the effective microscopic interactions that underlie a given thermodynamic system \cites{Ruppeiner:1979aa, Ruppeiner:1998aa}. We show that any manifold equipped with a degenerate Hessian metric whose potentials are extensive contains embedded Hessian submanifolds (Proposition \ref{prp:HessSubm}). The study of their scalar curvature represents a mathematical generalization of Ruppeiner geometry. Under this more general approach, scalar curvature appears to yield information about thermodynamic critical points, but the relationship between its sign and the nature of effective interactions at the microscopic level seems to be lost.

This paper is organized as follows. Section \ref{sec:edfrm} is devoted to motivating the definition of extensive differential forms on manifolds in a coordinate-free way. For this end, we review the very basic ideas of affine and radiant manifolds, upon which this definition lies.

Section \ref{sec:xHs} features the object of study of this paper: degenerate Hessian structures whose potentials are extensive up to an additive constant. We present global characterizations of such structures.

We show in Section \ref{sec:hsdhs} that any manifold endowed with a degenerate Hessian structure having extensive local potentials possesses Hessian submanifolds, which are the setting of Ruppeiner geometry. These submanifolds are embedded, and locally portrayed as level sets of certain distinguished coordinate functions that are the mathematical analogue of the ``natural variables'' of the entropy of a thermodynamic system.

We apply the ideas developed in the previous sections to the study of thermodynamic systems in Section \ref{sec:RG}. We show through three examples that the sign of the scalar curvature of Hessian submanifolds depends upon the choice of the submanifold. Besides, thermodynamic critical points seem to correspond to geometric singularities. We discuss the implications of these findings in relation to Ruppeiner's interaction hypothesis.

Finally, Section \ref{sec:concl} is dedicated to concluding remarks and perspectives.  
 
 \section{Extensive differential forms on radiant manifolds}\label{sec:edfrm}
In what follows, $M$ denotes $n$-dimensional smooth manifold, with $n\in\mathbb{N}$. All vector fields and differential forms (including functions) to which we refer are assumed to be smooth. Einstein's summation convention over repeated indices is used. 

We begin by briefly reviewing some of the geometric concepts that we shall use throughout the paper. Recall that an \emph{affine manifold} is a pair $(M,\bar\nabla)$, where $\bar\nabla$ is a symmetric, flat affine connection on $M$. Equivalently, an affine manifold may be described as one endowed with an atlas whose coordinate changes are locally affine transformations on certain affine space. In this paper, we call this atlas an \emph{affine structure} on $M$, and refer to the charts belonging thereto as \emph{affine charts}. Observe that the Christoffel symbols of $\bar\nabla$ vanish in a holonimic basis if and only if this basis is induced by an affine chart. 

Affine structures allow for a coordinate-independent definition of the Hessian of real functions. This is useful in the context of equilibrium thermodynamics since, as we explained before, the spaces of equilibrium states of thermodynamic systems are endowed with a metric tensor whose components are given by the Hessian matrix of a thermodynamic potential in certain coordinate chart. An important feature of the potentials of these structures is that they are  \textit{extensive functions}. Their geometric description requires a further structure, which we describe below.  

An affine manifold $(M,\bar\nabla)$ is said to be \emph{radiant} if it admits a global vector field $\rho$ such that 
\begin{equation}\label{eq:rad}
\bar\nabla_X\rho=X,
\end{equation} 
\noindent for every local vector field $X$. Regarded as a tangent-bundle-valued 1-form, Eq. \eqref{eq:rad} states that $\bar\nabla\rho$ must be the identity mapping. The existence of such a vector field, commonly referred to as \emph{radiant vector field}, is equivalent to the existence of an atlas on $M$ whose coordinate changes are locally linear transformations \cite{Goldman:aa}*{Prop. 4.17}. We shall refer to this atlas as a \emph{radiant structure} on $M$, and to its corresponding charts as \emph{radiant charts}. The triple formed by $(M,\bar\nabla,\rho)$ is called \emph{radiant manifold}. There is an analogy between the relationship of affine structures to flat connections and the relationship of radiant structures to radiant vector fields. Namely, a smooth chart $(U,(x^1,\ldots,x^n))$ on $M$ is radiant if and only if $\rho$ is written in this chart as an \textit{Euler vector field}, this is, $\rho|_U=x^i\partial_i$ (the symbol $\partial_i$ stands for $\partial/\partial x^i$ in the last equation and in every in-line expression henceforth, as long as there is no chance of confusion).

We now turn our attention to the concept of \emph{extensive function}. Recall that a function is said to be \emph{extensive} when it is a homogeneous degree-one function of the ``extensive variables'' of the system. Despite missing a precise mathematical definition of ``extensive variables'', we can define extensive functions on radiant manifolds by recalling that a function $f$ defined on an open set of the Euclidean space is a homogeneous first-order function if and only if it satisfies the Euler equation: $f=u^i\partial_i f$, where $u^1,\ldots,u^n$ are the cartesian coordinates thereon. The particular form of $\rho$ on radiant charts allows for a definition by analogy of extensive functions on a radiant manifold $(M,\Del,\rho)$: a local smooth function on $M$ is extensive if $\ed f (\rho)=f$. Upon observing that $\ed f(\rho)=\Lie{\rho}f$, we can extend this notion to differential forms as follows (\textit{cf.} Ref. \cite{Belgiorno:2002aa}).

\begin{dfn}\label{dfn:ext}
Let $k\in\N\cup\{0\}$. A local $k$-form $\omega$ defined on $M$ is said to be \emph{extensive} if 
\begin{equation}\label{eq:ext}
\Lie{\rho}\omega=\omega.
\end{equation}
\end{dfn}

We stress that the definition above is necessary for the study of the degenerate Hessian structures that appear in thermodynamics, since the metric potentials of these structures are extensive. In the next section, we shall translate this condition upon local potentials to a global one. Before paying attention to that matter, we provide a characterization of extensive $k$-forms in terms of the covariant derivative along $\rho$.

\begin{prp}\label{prp:extensivecovariant}
Let $k\in\N\cup\{0\}$. A local $k$-form $\omega$ is extensive if and only if
\begin{equation}\label{eq:extensivecovariant}
\cd{\rho}{\omega}=(1-k)\omega.
\end{equation}
\end{prp}

\begin{proof}
Let $X_1,X_2,\ldots,X_k$ be any local vector fields on $M$. Then $\cd{\rho}{\omega}(X_1\wedge\cdots\wedge X_k)=\ed(\omega(X_1\wedge\cdots\wedge X_k))(\rho)-\omega(\cd{\rho}X_1\wedge\cdots\wedge X_k)-\cdots-\omega(X_1\wedge\cdots\wedge\cd{\rho}X_k)=\Lie{\rho}\omega(X_1\wedge\cdots\wedge X_k)+\omega(\Lie{\rho}X_1\wedge\cdots\wedge X_k)+\cdots+\omega(X_1\wedge\cdots\wedge\Lie{\rho}X_k)-\omega(\cd{\rho}X_1\wedge\cdots\wedge X_k)-\cdots-\omega(X_1\wedge\cdots\wedge\cd{\rho}X_k)=\Lie{\rho}\omega(X_1\wedge\cdots\wedge X_k)+\omega((\Lie{\rho}X_1-\cd{\rho}X_1))\wedge\cdots\wedge X_k)+\cdots+\omega(X_1\wedge\cdots\wedge(\Lie{\rho}X_k-\cd{\rho}X_k))$.

Recall that $\Del$ is symmetric, which implies that for every $i\in\{1,\ldots,k\}$, $\Lie{\rho}X_i-\cd{\rho}X_i=-\cd{X_i}\rho$. The right-hand side of the last equation is in turn equal to $X_i$ (see Eq. \eqref{eq:rad}). We have therefore that $\cd{\rho}\omega=\Lie{\rho}\omega-k\omega$, whence the desired result follows.
\end{proof}

In particular, owing to the previous result, a local function on $M$ is extensive if and only if $\cd{\rho}f=f$, whereas a 1-form $\omega$ is extensive if and only if $\cd{\rho}\omega=0$. It is worth remarking that the idea of extensive differential form is applicable to any (covariant) $k$-tensor field on $M$. The previous result also holds in such case.

\section{Extensive Hessian structures}\label{sec:xHs}

In this section, we look for a global characterization of the Hessian structures whose potentials are extensive up to an additive constant (equivalently, whose differential is extensive), as is  the case in thermodynamics. It turns out that having such a potential renders a Hessian structure degenerate, as we show below. Henceforth, $(M,\Del,\rho)$ denotes a radiant manifold. For each local vector field $X$ on $M$ we denote by $X^\flat$ the 1-form given by $X^\flat(Y)=g(X,Y)$, for every local vector field $Y$.

\begin{thm}\label{thm:ex=>dg}
A Hessian structure $(\Del, g)$ on $M$ having local potentials with extensive differential around each point of $M$ is degenerate.
\end{thm}

\begin{proof}
Let $U$ be an open subset of $M$ and $\varPhi$ be a local potential for $g$ on $U$. Then $\rho^\flat|_U=\cd{\rho}\ed\varPhi$. From Proposition \ref{prp:extensivecovariant}, we have that if $\ed\varPhi$ is extensive, then
\begin{equation}\label{eq:GD}
\rho^\flat=0,
\end{equation}
which shows that $(\Del,g)$ is degenerate.
\end{proof}

The last result is a recipe to endow any radiant manifold with a degenerate Hessian structure, as the next example illustrates.

\begin{exm}\label{exm:1}
The plane $\R^2$ with the origin deleted, denoted by $\R^2_0$, is a radiant structure with $\Del$ given by the restriction hereto of the canonical flat connection of $\R^2$ and $\rho:=x\partial_x+y\partial_y$, where $x$ and $y$ denote the cartesian coordinates on the plane. 

The euclidean distance to the origin of any point $(x,y)\in\R^2_0$, denoted by $r$, is extensive. Indeed, $\ed r(\rho)=(x^2+y^2)^{-1/2}(x\ed x+y\ed y)(x\partial_x+y\partial_y)=r$. Hence, $r$ defines a degenerate Hessian structure on $\R^2_0$, provided that $\Del^2r$ is positive semidefinite, which can be verified straightforwardly.

If $\theta$ denotes the polar coordinate on a suitable open, simply-connected subset of $\R^2_0$, defined by $\theta:=\tan^{-1}(y/x)$, it can readily be seen that the degenerate Hessian metric $g$ that $r$ defines is locally $g=r(\ed\theta)^2$. 
\end{exm}

It is important to observe that if $(\Del, g)$ has extensive local potentials, it is not only degenerate, but the radiant vector field lies in the kernel of $\flat$, as Eq. \eqref{eq:GD} states (\textit{cf.} Example \ref{exm:1}). The latter is the coordinate-free version of the \emph{Gibbs-Duhem equation}, which is well known in thermodynamics. To make this evident, we take a radiant chart $(V,(x^1,\ldots,x^n))$ whose domain is contained in $U$. Then $\rho^\flat|_V=g_{ij}x^i\ed x^j=x^i\partial_i\partial_j\varPhi\ed x^j$. If we define $y_j:=\partial_j\varPhi$, for each $j\in\{1,\ldots,n\}$, we have that Eq. \eqref{eq:GD} is equivalent on $V$ to $x^i\ed y_i=0$, which is the familiar form of the Gibbs-Duhem equation. 

Owing to the local expression for $\rho^\flat$, we can readily observe that if Eq. \eqref{eq:GD} is satisfied, then any local potential for $g$ has extensive derivative. 

\begin{thm}\label{thm:dg=>ex}
If a degenerate Hessian structure $(\Del,g)$ on $M$ satisfies Eq. \eqref{eq:GD}, then the derivative of any potential of $g$ is extensive.
\end{thm}

The two results above imply for $(\Del,g)$ that having local potentials with extensive derivative is equivalent to having $\rho$ as a null vector. We have thus translated the former feature, which is local, to a global one represented by the Gibbs-Duhem equation. A Hessian structure that satisfies Eq. \eqref{eq:GD} has a particular behavior under the Lie derivative along $\rho$, as stems from the following, more general result.

\begin{prp}\label{prp:radiantCodazzi}
Let $g$ be a 2-tensor field on $M$ and suppose that the pair $(\Del,g)$ satisfies Eq. \eqref{eq:Codazzi}. Then\footnote{By $\Del\rho^\flat$ we mean the covariant derivative of $\rho^\flat$. Lowering the index of $\Del\rho$, which yields $g$, would be written as $({\Del\rho})^\flat$, to avoid confusion.}
\begin{equation}\label{eq:radiantCodazzi}
\Lie{\rho}g=g+\Del\rho^\flat.
\end{equation} 
\end{prp}

\begin{proof}
Let $U$ be any open subset of $M$ and $X,Y\in\mathfrak{X}(U)$. Then, $\Lie{\rho}g(X,Y)=\ed(g(X,Y))(\rho)-g(\Lie{\rho}X,Y)-g(X,\Lie{\rho}Y)$. Since $\Del$ is symmetric, the last two terms of the right-hand side of the last equation may be rewritten as $g(\Del_{\rho}X-\Del_X\rho,Y)$ and $g(X,\Del_\rho Y-\Del_Y\rho)$, respectively. Considering that $\Del\rho=\mathrm{id}$, we have that $\Lie{\rho}g(X,Y)=\ed(g(X,Y))(\rho)-g(\Del_\rho X,Y)-g(X,\Del_\rho Y)+2g(X,Y)=\Del_\rho g(X,Y)+2g(X,Y)$. Since $(\Del,g)$ satisfies Eq. \eqref{eq:Codazzi}, then $\Lie{\rho}g(X,Y)=\Del_X g(\rho, Y)+2g(X,Y)=\ed(g(\rho,Y))(X)-g(\Del_X\rho,Y)-g(\rho,\Del_XY)+2g(X,Y)=\ed(g(\rho,Y))(X)-g(\rho,\Del_XY)+g(X,Y)=\Del_X\rho^\flat(Y)+g(X,Y)$. Because the last equation holds for any local vector fields on $M$, Eq. \eqref{eq:radiantCodazzi} follows. 
\end{proof}

As a consequence, given a pair $(\Del,g)$ on $M$, we have that $\rho$ is conformal (with respect to $g$) if and only if $\rho^\flat$ is parallel. In particular, as we mentioned before, if $\rho^\flat=0$, then $g$ satisfies Eq. \eqref{eq:ext}, mimicking the behavior of extensive differential forms. This motivates the following definition.

\begin{dfn}
A degenerate Hessian structure $(\Del,g)$ on $M$ is said to be \emph{extensive} if it satisfies Eq. \eqref{eq:GD}.
\end{dfn} 

Theorems \ref{thm:ex=>dg} and \ref{thm:dg=>ex} imply that a Hessian structure is extensive if and only if its local potentials are extensive. Furthermore, from Proposition \ref{prp:radiantCodazzi} it follows that  $\rho$ is a conformal vector field of $g$, whenever $(\Del, g)$ is extensive. The converse is not true, however, as we make evident below.

\begin{exm}
Consider the Euclidean $n$-space $\R^n$ with its standard radiant structure, and let $\R^n_+$ denote the submanifold $\{p\in\R^n:u^1(p),\ldots,u^n(p)>0\}$, where $u^1,\ldots,u^n$ denote the cartesian coordinates on $\R^n$. We let $(\Del,\rho)$ stand for the radiant structure that $\R^n_+$ inherits from $\R^n$.

We define $\phi:\R^n_+\to\R$ by $\phi=\sum_{i=1}^nu^i(\ln\circ u^i-1)$. It can readily be seen that $\phi$ is a global potential for the metric $g:=\sum_{i=1}^n(\ed u^i)^2/u^i$, which is Riemannian. 

A straightforward computation yields $\rho^\flat=\ed(u^1+\ldots+u^n)$, which is parallel with respect to $\Del$. According to Theorem \ref{thm:dg=>ex}, this implies that $\rho$ is a conformal vector field, yet $(\Del,g)$ is not extensive (since it is non-degenerate).
\end{exm}

Extensive Hessian structures behave as expected under the covariant derivative along $\rho$ (see Proposition \ref{prp:extensivecovariant}). This follows from the next result.

\begin{prp}
Let $(\Del,g)$ be like in Proposition \ref{prp:radiantCodazzi}. Then,
\begin{equation}\label{eq:covariantCodazzi}
\cd{\rho}g=-g+\Del\rho^\flat.
\end{equation}
\end{prp}

\begin{proof}
Suppose that $X$ and $Y$ are any two local vector fields on $M$. Then $\cd{\rho}g(X,Y)=\cd{X}g(\rho,Y)=\ed(\rho^\flat(Y))(X)-g(X,Y)-\rho^\flat(\cd{X}Y)=\Del\rho^\flat(X,Y)-g(X,Y)$, as we wished to prove.
\end{proof}

So far we have presented a geometric framework that is particularly motivated by the so-called \emph{Ruppeiner geometry}. In the language of this paper, Ruppeiner geometry is the study of the scalar curvature of certain Hessian submanifolds of a manifold endowed with an extensive Hessian structure. In the next section, we shall show that any radiant manifold endowed with an extensive Hessian structure possesses Riemannian submanifolds, characterized by being those transversal to $\ker\flat$. We also prove the existence of Hessian submanifolds, and show that these are embedded. These submanifolds correspond to the setting of Ruppeiner geometry.

\section{Hessian submanifolds}\label{sec:hsdhs}

As we have mentioned throughout the paper, the Hessian submanifolds of the manifold of states of a thermodynamic system are of particular importance to physics. In this section, we show that their existence follows from the fact that spaces of equilibrium states are endowed with an extensive Hessian structure. In what follows, $(\Del,g)$ represents an extensive Hessian structure on $M$.

We begin by characterizing the Riemannian submanifolds of $M$: they are manifolds whose tangent space at each point does not contain any null vectors of $g$. To state this in precise terms, we show first that $\ker\flat$ defines an integrable distribution on $M$ (we remind the reader that a distribution is integrable if and only if it is involutive).

\begin{prp}
The distribution defined by $\ker\flat$ is involutive.
\end{prp}

\begin{proof}
Since $\Del$ is symmetric, for any local vector fields $X$, and $Y$ with common domain on $M$ one has that $[X,Y]^\flat=(\cd{X}Y)^\flat-(\cd{Y}X)^\flat$. 

Let $Z$ be a local vector field on $M$ sharing domain with $X$ and $Y$. Then $(\cd{X}Y)^\flat(Z)=g(\cd{X}Y,Z)=\ed(Y^\flat(Z))(X)-\cd{X}g(Y,Z)-Y^\flat(\cd{X}Z)=\cd{X}Y^\flat(Z)-\cd{X}g(Y,Z)$. Therefore, $[X,Y]^\flat(Z)=\cd{X}Y^\flat(Z)-\cd{X}g(Y,Z)-\cd{Y}X^\flat(Z)+\cd{Y}g(X,Z)$.

From Eq. \eqref{eq:Codazzi}, it follows that $[X,Y]^\flat=\cd{X}Y^\flat-\cd{Y}X^\flat$.  Thus, $X^\flat=Y^\flat=0$ implies $[X,Y]^\flat=0$. This means that the distribution defined by $\ker\flat$ is involutive, as we wished to prove.
\end{proof}

If $\imath:N\hookrightarrow M$ is a submanifold that is transversal to any integral submanifold of the distribution defined by $\ker\flat$, then it is Riemannian (with metric $\imath^*g$, where $\imath^*g$ denotes the pull-back of $g$ under the inclusion $\imath$). This follows upon observing that, since $g$ is positive semidefinite, $g(X,X)=0$ if and only if $X^\flat=0$. 

Although $\Del$ is flat, the corresponding induced connection on a submanifold $\imath$ of $M$ might not be so\footnote{Whenever there is no chance of confusion, we shall refer to a submanifold of $M$ by its inclusion.}. However, any manifold defined by a level set of codimension $\dim\ker\flat$ of coordinate transformations corresponding to a radiant chart is Hessian (this is, it is transversal to $\ker\flat$ and the pull-back connection is both symmetric and flat). Indeed, let $\varPhi$ be a potential defined on an open subset $U$ of $M$ and let $(U,(x^1,\ldots,x^n))$ be a radiant chart. Since $g$ is degenerate, $\det\flat=0$, which is translated locally to $\det(\partial_iy_j)=0$, where $y_j:=\partial_j\varPhi$, for every $j\in\{1,\ldots,n\}$. Therefore, $n-\dim\ker\flat$ functions of the set $\{y_1,\ldots,y_n\}$ are independent. Assume that $y_1,\ldots,y_{n-\dim\ker\flat}$ are independent. Then, the submanifold defined by $\imath^*\ed x^j=0$, for all $j\in\{n-\dim\ker\flat+1,\ldots,n\}$ is Riemannian, with metric $\imath^*(g|_U)$. This is true precisely because $\det(\imath^*\partial_iy_j)\neq0$  \cite{Torres-del-Castillo:1993aa}. We can readily verify that $\imath$ is also Hessian. In order to do so, recall that the covariant derivative of 1-forms corresponding to the pull-back connection, $\imath^*\Del$, is defined as $(\imath^*\Del)_X(\imath^*\alpha):=\imath^*(\Del_{\imath_*X}\alpha$), where $\imath_*$ denotes the derivative of $\imath$, for any local vector field $X$ defined on $N$ and any local 1-form $\alpha$ defined on $M$. For each $\alpha\in\{1,\ldots,n-\ker\flat\}$, let $z^\alpha:=\imath^*x^\alpha$.  Then, for every $\alpha,\beta\in\{1,\ldots,n-\dim\ker\flat\}$, $(\imath^*\Del)_{\partial/\partial z^\alpha}\ed z^\beta=\imath^*(\Del_{\imath_*\partial/\partial z^\alpha}\ed x^\beta)=\imath^*(\Del_{\partial/\partial x^\alpha}\ed x^\beta)=0$. We have thus constructed a Hessian submanifold of $M$.

Observe that the Hessian submanifolds of $M$ obtained by the procedure described above are also embedded. A Hessian coordinate chart is precisely a slice chart for $\imath$. It turns out that this is true for \emph{any} Hessian submanifold of $M$, as we now show.

\begin{prp}\label{prp:HessSubm}
Any Hessian submanifold $\imath$ of $M$ is embedded. Radiant coordinate charts are slice charts for $\imath$.
\end{prp}

\begin{proof}
Let $\imath:N\hookrightarrow M$ denote an $r$-dimensional Hessian submanifold of $M$, (notice that $r\leq\rank\flat$). Since $\imath^*\bar\nabla$ is affine, there exists a coordinate chart $(V,(z^1,\ldots,z^r))$ around any point $p\in N$, such that $(\operatorname{\imath^*\bar\nabla})_{\alpha}\ed z^\beta=0$, for all $\alpha,\beta\in\{1,\ldots,r\}$. 

If $(U,(x^1,\ldots,x^n))$ is a radiant chart around $\imath(p)$, then $\operatorname{(\imath^*\Del)}_{\alpha}\imath^*(\ed x^j)=\imath^*\left(\partial_\alpha\imath^*x^k\cd{k}\ed x^j\right),$ for all $\alpha\in\{1,\ldots,r\}$ and $j\in\{1,\ldots,n\}$. The last expression vanishes because $\Del$ is flat. 

On the other hand, $\operatorname{(\imath^*\Del)}_{\alpha}\imath^*(\ed x^j)=\partial_{\alpha}\partial_{\beta}\imath^*x^j\ed z^\beta.$ From the paragraph above, it follows that the latter 1-form must be zero. So, $\partial_\alpha\imath^*x^j=a^j_\alpha,$ with $a^j_\alpha\in\R$, for all $j\in\{1,\ldots,n\}$ and $\alpha\in\{1,\ldots,r\}$. Since $\imath$ is an immersion, $\rank\imath=r$, whence there exists an invertible $n\times n$ real matrix $(b_i^j)$ such that $b_i^ja^i_\alpha=0$, for each $j\in\{r+1,\ldots,n\}$ and $\alpha\in\{1,\ldots,r\}$. Therefore, the radiant chart $(U,(\tilde x^1,\ldots,\tilde x^n))$ given by $\tilde x^k:=b^k_ix^i$ is a slice coordinate chart for $\imath$ around $p$, which concludes the proof.
\end{proof}

It is important to point out that the coordinate charts whose level sets define locally Hessian submanifolds do not vanish. In other words, if $(U,(x^1,\ldots,x^n))$ is a radiant chart, $\imath^*x^{r+1}=\cdots=\imath^*x^n=0$ does not define a Hessian submanifold of $M$, because $\imath$ and $\ker\flat$ are not transversal to each other in that case (the tangent space to the image of $\imath$ and $\ker\flat$ share $\rho$ at each point).

As we mentioned before, the study of the Hessian submanifolds of $M$ is motivated by Ruppeiner geometry. The most important claim of this approach to thermodynamics is that the scalar curvature of the submanifold defined by constant volume yields information regarding microscopic interactions amongst the components of the system \cite{Ruppeiner:2010aa}. We know from the last result that such submanifold is not only Riemannian, but also a \emph{particular} Hessian submanifold of the manifold of states of the system. We argue that if scalar curvature yields any physical information, it should not rely on the Hessian submanifold that is chosen to compute it. This idea offers a generalization of Ruppeiner geometry that might be useful to retrieve physical information from geometry, for systems lacking a straightforward analogue of volume among their coordinates. We delve further into this idea in the next section, after dealing with a well-known result of Hessian geometry that is relevant in the context of thermodynamics. 

There is an important mathematical feature of thermodynamics that has not been addressed so far in this paper: the so-called \emph{Legendre invariance} of the description of a thermodynamic system. By the latter, it is understood that thermodynamic potentials restricted to hypersurfaces (submanifolds of codimension 1) of the manifold of states of a system are determined up to a Legendre transform. We stress two facts about the equivalence of thermodynamic potentials under Legendre transforms. First, this equivalence is only valid for potentials restricted to hypersurfaces (or submanifolds of codimension greater than one). Otherwise, the potential $\varPhi=0$ would be equivalent to any other potential, rendering thermodynamics trivial. Second, \emph{partial} Legendre transforms lack a coordinate-free description. Therefore, they are not treated in this paper. In contrast, \emph{total} Legendre transforms do possess a familiar coordinate-free approach in the setting of Hessian geometry, which we review below.

Like any Hessian manifold, a  Hessian submanifold $\imath:N\hookrightarrow M$ admits a \emph{dual Hessian structure}. Namely, if $\nabla$ denotes the Levi-Civita connection of $\imath^*g$, then $\bar\nabla^*:=2\nabla-\imath^*\bar\nabla$ is a flat connection which together with $\imath^*g$ forms a Hessian structure on $N$, called the \emph{dual Hessian structure} of $(\imath^*g,\imath^*\bar\nabla)$ \cite{Shima:2007aa}. Let $\varPhi$ be a local potential of $(\Del, g)$ sharing domain with a radiant chart $(U,(x^1,\ldots,x^n))$, which is also a slice chart for $\imath$ (this is, $\imath^*\ed x^j=0,$ for all $j\in\{\dim N+1,\ldots,n\}$). We denote by $z^\alpha$ the coordinate functions $\imath^*x^\alpha$, for all $\alpha\in\{1,\ldots,\dim N\}$, and define $y_\alpha:=\partial_\alpha\varPhi$. It can readily be seen that $(\imath^{-1}(U),(y_1,\ldots,y_{\dim N}))$ is an affine coordinate chart for $\Del^*$, \textit{i. e.}, the Christoffel symbols of $\Del^*$ vanish on the holonomic basis induced thereby \cite{Shima:2007aa}. In thermodynamics, the functions $(y_1,\ldots,y_{\dim N})$ are called the \emph{natural variables} of $\varPhi^*$.  

Since $\varPhi$ is a local potential of $(\Del, g)$ on $U$, then $\imath^*\varPhi$ is a potential for $(\imath^*\Del,\imath^*g)$, which shares domain with $\varPhi^*$. These two functions are related to each other through a \emph{total} Legendre transform. To be specific, $\varPhi^*=z^\alpha\partial_\alpha\imath^*\varPhi-\imath^*\varPhi.$ Observe that, since $\varPhi$ is extensive, the \emph {dual potential} $\varPhi^*$ may also be written as $\varPhi^*=-\imath^*\left(x^A\partial_A\varPhi\right),$ where $A$ runs through $\{\dim N+1,\ldots,n\}$. The fact that both $(\imath^*\bar\nabla,\imath^*g)$ and $(\bar\nabla^*,\imath^*g)$ form a Hessian structure may be translated as Ruppeiner geometry being invariant under \textit{total} Legendre transforms. 

\section{A generalization of Ruppeiner Geometry}\label{sec:RG}

We claimed before that the Hessian submanifolds of radiant manifolds endowed with extensive Hessian structures are of particular importance to physics. The reason, as we have pointed out, is that, in some particular cases, their scalar curvature seems to yield valuable information about the system, related to phase transitions and critical phenomena. The approach that we have briefly described, referred to herein as Ruppeiner geometry, considers only one of the many Hessian submanifolds that the manifold of states of a system possesses. This ``bias'' is of course motivated by physical reasons, and goes well beyond computational convenience. Because of the physical relevance of this approach to thermodynamics, we seek to expand it on purely mathematical grounds. In order to do this, we consider in this section three particular thermodynamic systems.

The first system we deal with is the ideal gas. As is well known, this system lacks (thermodynamic) critical points, which is geometrically translated to a flat geometry of the manifold defined by constant volume. However, as we show below, not every Hessian submanifold of its manifold of equilibrium states is flat. Yet, none of the latter has states at which its scalar curvature goes to infinity (which, under Ruppeiner's approach, are to be identified with thermodynamic critical points).

The second system under our consideration is the model of an ideal paramagnetic system, presented by Callen \cite{Callen:1985aa} mainly as a toy model. Surprisingly, under our approach, this system does behave like an ideal one, in the sense that all its Hessian submanifolds lack critical points, like the ideal gas. 

The last system that we analyze is the Kerr-Newman black hole family. Our study is performed under the approach that we have presented in this paper (see also \cite{Garcia-Ariza:2014aa}), which does not coincide with the standard geometric approaches to black hole thermodynamics. This system does possess a critical point, which occurs at absolute zero: precisely the states at which black holes undergo \emph{a transition} from being singularities with horizons into naked ones. 

In the three examples that we present here, it is possible to construct Hessian submanifolds of the manifold of states whose scalar curvature attains any sign (or vanishes). This disagrees \textit{a priori} with Ruppeiner's interpretation of the scalar curvature in thermodynamics. 

\subsection{The ideal gas}

Let $M_{\text{ig}}$ denote the 3-dimensional space of equilibrium states of a simple ideal gas. A global coordinate chart for this manifold is given by $(M_{\text{ig}},(U,V,N))$, where $V$ represents the volume of the system, and $U$ and $N$ its internal energy and number of particles, respectively. 

Because $M_\text{ig}$ is endowed with a global chart, we can globally define a radiant structure $(\Del_\text{ig},\rho_\text{ig})$ hereon by means of the coordinate expressions of the connection and the vector, respectively. Namely, we let $\rho_\text{ig}:=U\partial_U+V\partial_V+N\partial_N$ and $\Del$ be such that its Christoffel symbols with respect to the frame $\{\partial_U,\partial_V,\partial_N\}$ vanish.

The entropy of this system, which is a global potential for the Hessian structure $(\Del_\text{ig},g_\text{ig})$ corresponding to this manifold is given by
\begin{equation}\label{eq:ig}
S_\text{ig}=N\left\{A \log\left[\frac{V}{V_0}\left(\frac{N}{N_{0}}\right)^{-C - 1}
\left(\frac{U}{U_{0}}\right)^{C}\right]\right\} + S_0,
\end{equation}
\noindent where $A, N_0, U_0, V_0$, and $C$ are positive constants, and $S_0$ is another constant representing the entropy of the ideal gas at a given reference state. The fact that Eq. \eqref{eq:ig} provides a global potential for $g_\text{ig}$ implies that for all $p\in M_\text{ig}$, $U(p),V(p),N(p)>0$. 

It can readily be seen that $\ed S_\text{ig}$ is extensive and $\dim\ker\flat_\text{ig}=1$. Thus, $g_\text{ig}$ is extensive and $\ker\flat_\text{ig}$ is spanned by $\rho$. Hence, the dimension of any Hessian submanifold of $M_\text{ig}$ is at most 2. 

Let $\imath$ be a 2-dimensional Hessian submanifold of $M_{\text{ig}}$. Then, $\imath$ is locally defined by $\imath^*(aU+bV+cN)=K$, for some real numbers $a,b,c$, and $K$, fixed by the conditions $\det(\imath^*g_\text{ig})\neq0$ and $K\neq 0$ (\textit{cf.} Proposition \ref{prp:HessSubm}). Owing to the last condition, we may assume that $b\neq0$. Then, locally, $\imath^*V=K_1U+K_2N+K_3$, with $K_1,K_2,K_3\in\mathbb{R}$, and $K_3\neq0$. A straightforward computation shows that $ \det(\imath^*g_\text{ig})=(A^{2} K_{3}^{2} C)/(U^2V^2),$ which is nonzero, as expected.

The scalar curvature of $\imath^*g_\text{ig}$ is given by
\begin{equation}\label{eq:Rig}
R_\text{ig}={-\frac{ab(C+1)UV-acNU+bcCNV}{ACN(aU+bV+cN)^2}}.
\end{equation}
\noindent Observe that $R_\text{ig}$ contains no singularities. Furthermore, the curvature of the manifolds given by $U=\text{const.}$ ($b=c=0$), $V=\text{const.}$ ($a=c=0$), and $N=\text{const.}$ ($a=b=0$), respectively, is zero. 

We may choose a different set of coordinates that yields a Hessian submanifold with nonvanishing scalar curvature. For instance, consider the global radiant chart given by $(M_\text{ig},(U+N,V,N))$. Then $\imath$ defined by $\imath^*(U+N)=\text{const.}$ yields a Hessian submanifold of $M_\text{ig}$, whose scalar curvature is $U/[AC(U+N)^2]$, which is not even constant (see Eq. \eqref{eq:Rig}). 

The paragraph above raises the following question: if the sign of scalar curvature has any physical relevance, how can we know \textit{a priori} what Hessian manifold to choose to compute it? We have just shown that a wrong choice might lead to wrong conclusions about critical behavior and microscopic interactions within the system, assuming that the sign of scalar curvature does possess any physical information. We shall observe a similar behavior of the scalar curvature in the next example.

\subsection{Ideal paramagnetic solid}
We now analyze the geometry of the Hessian submanifolds of the manifold of states corresponding to a toy model of a simple ideal paramagnetic system. The fundamental equation of such a system in terms of its internal energy $U$, magnetic moment $I$, and number of particles $N$ is given by \cite{Callen:1985aa}*{p. 83}
\begin{equation}\label{eq:fip}
S_\text{ip}=\ -A N {\left[\frac{I^{2}}{I_{0}
N^{2}} - \log\left(\frac{U}{A N T_{0}}\right)\right]},
\end{equation}
\noindent where $A$, $T_0$ and $I_0$ are positive constants, and $U(p),N(p)>0$ for all states $p$ belonging to its manifold of equilibrium states, $M_\text{ip}$. We let $\Del_\text{ip}$ denote the flat connection corresponding to the radiant structure defined by the global chart $(M_\text{ip},(U,I,N))$. The tensor $g_\text{ip}:=-\Del^2S_\text{ip}$ is positive semi-definite, as can readily be verified. Since $\ed S_\text{ip}$ is extensive, $g_\text{ip}$ determines, together with $\Del_\text{ip}$, an extensive Hessian structure on $M_\text{ip}$.

Like in the example above, $\ker\flat_\text{ip}$ is spanned by $\rho$, whence any 2-dimensional Hessian submanifold $\imath$ of $M_\text{ip}$ may be assumed to be defined by $\imath^*I(p)=K_1\imath^*U(p)+K_2\imath^*N(p)+K_3,$with $K_3\neq0$. This is true because a linear combination of $(U,I,N)$ is a slice coordinate chart for $\imath$, and thus $\imath^*(aU+bI+cN)(p)=K$, for some constants $a,b,c,K\in\mathbb{R}$, with $K\neq0$, and $b\neq0$, which may be assumed with no loss of generality.

The determinant of the metric tensor $\imath^*g$ is $\det(\imath^*g)=(2 \, A^{2} K_{3}^{2})/(I_{0} N^{2} U^{2}),$ which is nonvanishing.

The scalar curvature of $\imath^*g_\text{ip}$ is
\begin{equation}\label{eq:Rip}
R_\text{ip}={\frac{2abIU+2acNU-b^2I_0N^2}{2AN(aU+bI+cN)^2}}.
\end{equation}
\noindent The manifolds defined by constant $U$ and constant $N$ have both vanishing scalar curvature. Yet, $I=\text{const.}$ has negative, nonconstant curvature. Again, the sign of the scalar curvature of two different, mathematically equivalent manifolds, yields inconsistent information, according to Ruppeiner's interaction hypothesis.

\subsection{The Kerr-Newman black hole family}
As a last example, we consider the Kerr-Newman black hole family. The manifold of equilibrium states of this system, $M_{\text{KN}}$ is also 3-dimensional, since black holes belonging to this family are characterized by their mass $M$, the magnitude of their angular momentum $L$, and their charge $Q$ \cite{Davies:1977aa}. As we mentioned before, we shall study the scalar curvature of the Hessian submanifolds of $M_{\text{KN}}$ following a nonstandard approach (see \cite{Garcia-Ariza:2014aa}). In rough terms, we demand that $M_{\text{KN}}$ has the geometric structure that we have presented in this paper. This means that the entropy of this system \emph{has} to be extensive (up to an additive constant). To achieve this, we consider the radiant structure $(\Del_\text{KN},\rho_\text{KN})$ defined by $\rho_\text{KN}:=x\partial_x+L\partial_L+y\partial_y$, where $x:=M^2$ and $y:=Q^2/2$, and $\Del_\text{KN}$ such that its Christoffel symbols with respect to $\{\partial_x, \partial_L,\partial_y\}$ vanish (compare to Ref. \cite{Ruppeiner:2008aa}; see also Ref. \cite{Aman:2015aa}). 

We asserted above that $x, L$ and $y$ are globally defined. This is indeed the case, because our study of Hessian submanifolds restricts our attention to nonvanishing linear functions of a given radiant chart. This means that $x$, $L$, and $y$ are considered to be nonzero, which implies that we are studying only positively (or negatively) charged black holes. In other words, we are assuming that for every $p\in M_\text{KN}$, (either) $Q(p)>0$ (or $Q(p)<0$).

In terms of the global coordinates defined above, the entropy for this black hole family is written as \cite{Davies:1978aa}
\begin{equation}\label{eq:Smarr}
S_\text{KN}=\frac{1}{4}\left[x\left(1+\sqrt{1-\frac{2y}{x}-\frac{L^2}{x^2}}\right)-y\right],
\end{equation}

\noindent which is defined only for states $p\in M_\text{KN}$ satisfying $x^2(p)>(2xy+L^2)(p)$, called \textit{nonextremal black holes} (massless black holes are thereby excluded). Observe furthermore that the entropy of positively- and negatively-charged black holes is the same, which justifies our paying attention only to positively-charged black holes. 

The choice of the radiant structure $(\Del_\text{KN},\rho_\text{KN})$ renders $g_\text{KN}:=-\Del_\text{KN}^2S_\text{KN}$ together with $\Del_\text{KN}$ an extensive Hessian structure. Hence, if $\imath$ is any Hessian submanifold of $M_{\text{KN}}$, then $\imath$ is locally defined by $\imath^*(ax+bL+cy)=K$, where $a,b,c$, and $K$ are constants. For convenience, we assume in this case that $c\neq0$, which yields $\imath^*y=K_1x+K_2L+K_3$, with $K_3\neq0$. It can readily be verified that $\det(\imath^*g_\text{KN})=K_3^2/\left[4x^2\left(1-2y/x-L^2/x^2\right)\right]^2.$  Thus, $\imath^*g_\text{KN}$ defines a metric on the tangent space to $M_\text{KN}$ at nonextremal black holes. However, as follows from the last equation, black holes at absolute zero (this is, points in $M_\text{KN}$ for which $1-2y/x-L^2/x^2$ vanishes) are singularities of the metric. To show that this is the case, we determine the scalar curvature of $\imath$. Indeed, a straightforward computation yields
\begin{equation}\label{eq:RKN}
R_\text{KN}=\frac{4x^2\left(1-\frac{2y}{x}-\frac{L^2}{x^2}\right)(b^2+c^2+2ac)+(ax+bL+cy)^2}{x(ax+bL+cy)^2\sqrt{1-\frac{2y}{x}-\frac{L^2}{x^2}}}.
\end{equation}
\noindent In purely geometric grounds, black holes at absolute zero are analogous to the two-phase states of van der Waals fluids \cite{Santoro:2005ab}. This analogy is strengthened by the fact that, when reaching absolute zero (becoming extremal), a black hole undergoes a \emph{transition} from a singularity with horizon into a naked one, resembling the qualitative change van der Waals fluids suffer when undergoing phase transitions. Excluding massless black holes, the only singularity that this generalization of Ruppeiner geometry exhibits is given by extremal black holes (compare to \cite{Ruppeiner:2008aa}). 

By setting in Eq. \eqref{eq:RKN} $a=b=0$, and $a=c=0$ , respectively, we obtain the same result we previously reported in Ref. \cite{Garcia-Ariza:2014aa} for manifolds given by constant angular momentum and constant charge. Observe that, unlike the two previous examples, we cannot construct a 2-dimensional Hessian submanifold of $M_\text{KN}$ with vanishing curvature. This is true because, otherwise, the values of $x$, $y$, and $L$ would be further restricted to satisfy $x^2-2xy-L^2=\text{const.}$, provided that $ax+bL+cy$ is constant.

\section{Concluding remarks}\label{sec:concl}
In this paper, we have portrayed Ruppeiner geometry as a particular case of a more general structure, which we obtain by means of putting together a radiant and a Hessian structure. As we showed, the former is useful to deal with extensive differential forms in a coordinate-free fashion. Demanding that the local potentials of the Hessian structure in question be extensive up to an additive constant was translated to the global condition of requiring that the radiant vector field be a conformal vector of the Hessian metric. This, as we have seen, is equivalent to the Gibbs-Duhem equation, which in geometric terms means that the radiant vector field is a null vector of the Hessian metric.

We point out that under the formalism we have presented, any other nonequivalent Hessian approach to thermodynamics may be treated \cite{Bravetti:2012ac}. This is, the metric potential of the metric must not necessarily be the entropy of the system. However, as is well known, the results obtained using two different potentials may vary significantly, which poses a challenge to generalize even further the ideas we have presented here, so that the dependence of results on the potentials is well understood, and a criterion for the choice of a potential may be proposed.

In the examples we have analyzed in the last section, we have shown that the scalar curvature of any Hessian submanifold of a manifold of equilibrium states, in general, seems to yield information about thermodynamic critical points. Despite the results that the procedure we have presented are encouraging, we must remark that the relationship between critical points and singularities, if any, still lacks a formal proof. On the other hand, much cannot be said about the relationship between the sign of scalar curvature and effective microscopic interactions. The reason is that the sign of the scalar curvature depends sensibly upon the choice of the Hessian submanifold in which it is being computed. There is \textit{a priori} no principle to define, from a mathematical point of view, what submanifold is preferred to perform this computation. We believe this problem is worth addressing, in order to understand from a geometric point of view the validity of the results that Ruppeiner geometry yields, on one hand, and to be able to export the same ideas to black hole thermodynamics rigorously, on the other.

There is one more issue to be addressed in the future: the choice of a radiant structure for a given thermodynamic system. In other words, how do we know with respect to what set of coordinates must the Hessian of the entropy be computed in Ruppeiner geometry? This is apparently straightforward in familiar cases, like hydrostatic systems. Nevertheless, this is not so for black holes. We have based our choice, which differs from the standard one, on the general theory presented in this paper. Yet, it is not clear from the outset that black hole thermodynamics should or should not share the same mathematical structure of other common thermodynamic systems.

It is important to mention that a coordinate-free approach to extensive differential forms does not require a radiant structure. This subject is currently under study and shall be reported elsewhere.

\section*{Acknowledgements}
The author thanks Professors G. F. Torres del Castillo and M. Montesinos for the encouraging discussions and the advice that helped to shape this paper. The author is also grateful to B. D\'iaz for his comments and suggestions on the manuscript, and to I. Rubalcava-Garc\'ia for her aid in the diffusion of this work. This work was partially supported by CONACyT, M\'exico, grant number 374393.

\bibliography{/Users/magarciaariza/Dropbox/scilib/mybib}
\end{document}